\documentstyle[12pt,epsfig]{article}
\begin{document}
\noindent
\begin{center}
{\Large {\bf Non-minimal Gravitational Coupling of \\
Phantom and Big Rip Singularity }}\\ \vspace{2cm}
 ${\bf Yousef~Bisabr}$\footnote{e-mail:~y-bisabr@srttu.edu.}\\
\vspace{.5cm} {\small{Department of Physics, Shahid Rajaee Teacher
Training University,
Lavizan, Tehran 16788, Iran}}\\
\end{center}
\vspace{1cm}
\begin{abstract}
We consider a non-minimal coupling of a perfect fluid matter system
with geometry, which the coupling function is taken to be an
arbitrary function of the Ricci scalar.  Due to such a coupling, the
matter stress tensor is no longer conserved and there is an energy
transfer between the two components.  By solving the conservation
equation and applying the second law of thermodynamics, we show that
direction of the energy transfer depends on the equation of state of
the matter fluid. In particular, a phantom fluid should loose energy
with expansion of the universe. This energy reduction can avoid the
universe to end with a cosmic doomsday.

\end{abstract}
~~~~~~~PACS Numbers: 04.50.Kd, 04.20.Cv, 95.36.+x \vspace{3cm}

Observations of distant supernovae have given strong evidences for
accelerating expansion of the universe \cite{ac}.  One approach to
describe this phenomena is to invoke a new matter component, usually
referred to as dark energy, described by an equation of state
parameter $w\equiv p/\rho$, namely, the ratio of the
homogeneous dark energy pressure over the energy density.  For a
cosmic speed up, one should have $w<-1/3$ which corresponds to
an exotic matter component which violates the strong energy
condition.  Analysis of the data sets reveals that the equation of
state parameter $w$ actually lies in a narrow region around the
line $w=-1$ and may even be smaller than this barrier which
then violates weak energy condition \cite{h}.  In this case the
matter component is referred to as phantom dark energy
\cite{caldwel}.  One of the important problems with this phantom
fluid is that its energy density grows with expansion of the
universe so that it blows up at a finite time in the future, the
so-called Big Rip singularity
\cite{cald}\footnote{For a recent discussion, see \cite{cap}.}.\\
It is shown \cite{no} that a replacement of the gravity sector with
power law $f(R)$ models may offer a scenario to avoid this future
singularity.  However, recent analysis precludes the possibility of
regarding power law $f(R)$ models as a viable candidate for
generalizing  the gravity sector \cite{bis}.  As a more promising
approach, one may consider an interaction between the phantom fluid
and dark matter which leads to reduction of energy density of the
former with expansion of the universe in certain conditions. It is
shown that the Big Rip singularity can be actually prevented in
certain regions of parameters space in these interacting models
\cite{int} \cite{amen}.  However, comparing these results with SNIa
data reveals that those regions of parameters space are unlikely at
more than $99$-percent
confidence level \cite{amen}.\\
In this note, we consider an interacting phantom model in which the
phantom fluid has a non-minimal interaction with geometry via an
arbitrary function of the Ricci scalar. The non-minimal coupling for
ordinary matter has been already introduced in order to explain the
flatness of the rotation curves of galaxies \cite{3a}. As a
generalization of $f(R)$ gravity models, the action functional can
be written as
\begin{equation}
S=\int d^4x \sqrt{-g} \{\frac{1}{2}f_1(R)+[1+\lambda f_2(R)]L_m\}
\label{1}\end{equation} where $f_1(R)$ and $f_2(R)$ are arbitrary
functions of the Ricci scalar $R$ and $L_m$ is the Lagrangian
density corresponding to matter systems.  The parameter $\lambda$
characterizes the strength of the non-minimal coupling of $f_2(R)$
with matter Lagrangian. When $\lambda=0$, there is no such an
anomalous gravitational coupling of matter systems.  In this case,
the choice $f_1(R)=2 \kappa R$ with $\kappa=(16\pi G)^{-1}$ gives
the standard Einstein-Hilbert action while a nonlinear $f_1(R)$
function corresponds to the usual $f(R)$ modified Gravity.  In order
to make our analysis less complicated and since we are only
interested in effects of the non-minimal coupling, we will
take $f_1(R)$ as linear and set $f_1(R) = 2\kappa R$.  In this case, the action (\ref{1}) is similar
to $f(R, L_m)$ modified gravity models proposed in \cite{new1} with $f(R, L_m) = \kappa R + (1+\lambda f_2(R)) L_m$.  There are also other generalizations
recently proposed in the literature such as $f(R, T)$  \cite{new2} and $f(R, T, R_{\mu\nu}T^{\mu\nu} )$ models \cite{new3}. \\
Varying the action with respect to the metric $g_{\mu\nu}$ yields
the field equations, given by,
$$
2(\kappa+\lambda f'_2(R)L_m)R_{\mu\nu}-\kappa
g_{\mu\nu}R=2\lambda(\nabla_{\mu}\nabla_{\nu}-g_{\mu\nu}\Box)
f'_2(R)L_m
$$
\begin{equation}
~~~~~~~~~~~~~~~~~~~~~~~~~~~+[1+\lambda f_2(R)]T^m_{\mu\nu}
\label{2}\end{equation}
where the prime represents the derivative with respect to the scalar curvature.  The matter energy-momentum tensor is defined as
\begin{equation}
T^m_{\mu\nu}=\frac{-2}{\sqrt{-g}}\frac{\delta(\sqrt{-g}L_m)}{\delta{g^{\mu\nu}}}
\label{3}\end{equation} which due to the explicit coupling of matter
with Ricci scalar satisfies
\begin{equation}
\nabla^{\mu}T^m_{\mu\nu}=\frac{\lambda f'_2(R)}{1+\lambda f_2(R)}(L_mg_{\mu\nu}-T^m_{\mu\nu})\nabla^{\mu} R
\label{4}\end{equation}
The coupling between matter systems and the higher
derivative curvature terms describes transferring  energy and momentum between matter and geometry beyond the
usual one already existed in curved spaces.  Details of this energy exchange depend on the explicit form
of the matter Lagrangian density $L_m$.  Here we consider a perfect fluid energy-momentum tensor as a matter system
\begin{equation}
T_{\mu\nu}=(\rho_m+p_m)u_{\mu}u_{\nu}+p_mg_{\mu\nu}
\label{b1}\end{equation} where $\rho_m$ and $p_m$ are energy density
and pressure, respectively. The four-velocity of the fluid is
denoted by $u_{\mu}$.  A phantom fluid is a matter system with
equation of state $w=p_m/\rho_m<-1$.  There are different
choices for the perfect fluid Lagrangian density which all of them
leads to the same energy-momentum tensor and field equations in the
context of general relativity \cite{2} \cite{3}.  The two Lagrangian
densities that have been widely used in the literature are $L_m=p_m$
and $L_m=-\rho_m$ \cite {3a} \cite{a3a} \cite{4} \cite{5}.  For a
perfect fluid that does not couple explicitly to the curvature
(i.e., for $\lambda = 0$), the two Lagrangian densities $L_m =p_m$
and $L_m=-\rho_m$ are perfectly equivalent, as discussed in \cite{4}
\cite{5}. However, in the model presented here the expression of
$L_m$ enters explicitly the field equations
and all results strongly depend on the choice of $L_m$.  In fact, it is shown that there is a strong debate about equivalency of different expressions of the Lagrangian density of a coupled perfect fluid ($\lambda \neq 0$) \cite{6}. Here we will take $L_m=p_m$ as the Lagrangian density of the matter fluid. \\
We project (\ref{4}) onto
the direction of the four-velocity which satisfies the conditions $u_{\mu}u^{\mu}=-1$ and
$u^{\nu}\nabla_{\mu}u_{\nu}=0$.  We also assume that $p_m=w \rho_m$ with $\omega$ being a constant equation of state parameter.
Then, contracting (\ref{4}) with $u^{\mu}$ gives the conservation equation
\begin{equation}
u^{\mu}\nabla_{\mu}\rho_m+(\omega+1)\rho_m \nabla_{\mu}u^{\mu}=-\frac{\lambda f'_2(R)}{1+\lambda f_2(R)}(L_m+\rho_m)u_{\nu}
\nabla^{\nu}R
\label{b2}\end{equation}
We use Friedmann-Robertson-Walker metric given by the line element
\begin{equation}
ds^2=-dt^2+a^2(t)(\frac{dr^2}{\sqrt{1-kr^2}}+d\Omega^2)
\label{b3}\end{equation}
where $a(t)$ is the scale factor.  Homogeneity and isotropy of the universe imply that $u^{\mu}=(1,0,0,0)$ and $\Gamma^{1}_{10}=\Gamma^{2}_{20}=
\Gamma^{3}_{30}=H$ where $H=\frac{\dot{a}}{a}$ is the Hubble parameter and an overdot indicates differentiation with respect to the cosmic
time $t$.  The expression (\ref{b2}) is then reduced to
\begin{equation}
\dot{\rho}_m+3H(w+1)\rho_m=-\frac{\lambda f'_2(R)}{1+\lambda
f_2(R)}(L_m+\rho_m)\dot{R} \label{b4}\end{equation} In general, the
fluid energy is not conserved due to the explicit fluid-curvature
coupling.  The right hand side of (\ref{b4}) acts as a source term
describing the energy transfer per unit time and per unit
volume\footnote{This is a general statement and there are some
situations that in spite of such a coupling the right hand side of
(\ref{b4}) vanishes.  For a discussion, see \cite{bis1}.}.\\ We now
project (\ref{4}) onto the direction normal to the four-velocity by
the use of the projection operator
$h_{\mu\nu}=u_{\mu}u_{\nu}+g_{\mu\nu}$.  This results in
$$
h^{\mu\alpha}\nabla^{\nu}T_{\mu\nu}=(w+1)\rho_m u_{\nu}\nabla^{\nu}u^{\alpha}+\nabla^{\alpha}p_m+u^{\mu}u^{\alpha}\nabla_{\mu}p_m
$$
\begin{equation}
~=\frac{\lambda f'_2(R)}{1+\lambda f_2(R)}(L_m-p_m)h^{\alpha \nu}\nabla_{\nu}R
\label{b5}\end{equation}
This is equivalent to
\begin{equation}
u_{\nu}\nabla^{\nu}u^{\alpha}=\frac{du^{\alpha}}{d\tau}+\Gamma^{\alpha}_{\beta\gamma}u^{\beta}u^{\gamma}=f^{\alpha}
\label{b6}\end{equation}
with
\begin{equation}
f^{\alpha}=\frac{1}{(w+1)\rho_m}[\frac{\lambda f'_2(R)}{1+\lambda f_2(R)}(L_m-p_m)\nabla_{\nu}R+\nabla_{\nu}P]h^{\alpha\nu}
\label{b7}\end{equation}
This is an additional force exerted on a fluid element implying a non-geodesic motion.  Notice that
since $h^{\alpha\nu}u_\alpha=0$, we have $f^{\alpha}u_{\alpha}=0$ and the additional force is orthogonal to
the four-velocity.  This is consistent with the usual
interpretation of the four-force, according to which only the
component of the force orthogonal to the
particles four-velocity can influence their trajectory.\\
The additional force due to the non-minimal coupling should be
attributed to the first term.  The second term proportional to the
pressure gradient does not exhibit a new effect and is the usual
term that appears in equations of motion of a relativistic fluid. In
our choice, $L_m=p_m$, the first term on the right hand side of
(\ref{b7}) vanishes implying that fluid elements follow geodesics of
the background metric and there is no additional force.  In this
case, matter is still non-conserved and the equation (\ref{b4})
takes the form
\begin{equation}
\dot{\rho}_m+3H(\rho_m+p_m)=-\frac{\lambda f'_2(R)}{1+\lambda
f_2(R)}(w+1)\rho_m \dot{R} \label{b8}\end{equation} To make a
closer look at this equation, we assume a power law expansion for
the scale factor $a(t)=a_0t^m$ and we adopt $f_2(R)=\alpha R^n$ with
$\alpha$, $n$, $a_0$ and $m$ being constant parameters.
Putting these forms into
(\ref{b8}), gives
\begin{equation}
\dot{\rho}_m+3H(\rho_m+p_m)=-\frac{\lambda n\alpha R^{n-1}}{1+\lambda \alpha R^n}(w+1)\rho_m \dot{R}
\label{b9}\end{equation}
To solve this equation, we consider two different cases in the following:\\\\
{\bf 1. The case} $\lambda \alpha R^{n}<<1$, in which (\ref{b9}) takes the form
\begin{equation}
\dot{\rho}_m+3H(\rho_m+p_m)=-nx(w+1)\rho_m \frac{\dot{R}}{R}
\label{b15}\end{equation} where $x=\lambda \alpha R^n$. we have
$$
H=mt^{-1}
$$
\begin{equation}
R=6(\dot{H}+2H^2)=6m(2m-1)t^{-2} \label{b11}\end{equation}
$$
\frac{\dot{R}}{R}=-2\frac{H}{m}
$$
 By substituting these results into
(\ref{b15}), we obtain the relation
\begin{equation}
\dot{\rho}_m+3H(1-\frac{2n}{3m}x)(\rho_m+p_m)=0
\label{b16}\end{equation}
Since $x<<1$, when $\frac{2n}{3m}$ remains of order of unity, we have $(1-\frac{2n}{3m}x)\approx 1$.  Thus
\begin{equation}
\dot{\rho}_m+3H(\rho_m+p_m)\approx0
\label{b17}\end{equation}
which gives evolution of matter energy density as the standard one
\begin{equation}
\rho_m \approx \bar{\rho_{0}}a^{-3(w+1)}
\label{b18}\end{equation}
with $\bar{\rho}_0$ being an integration constant.  In this case matter is conserved and there is no creation or annihilation.\\\\
{\bf 2. The case} $\lambda \alpha R^{n}>>1$, in which (\ref{b9}) reduces to
\begin{equation}
\dot{\rho}_m+3H(\rho_m+p_m)=-n(w+1)\rho_m \frac{\dot{R}}{R}
\label{b10}\end{equation} Combining this result with (\ref{b11})
gives
\begin{equation}
\dot{\rho}_m+3\gamma H\rho_m=0
\label{b12}\end{equation}
where $\gamma=(1-\frac{2n}{3m})(w+1)$.  This is a simple differential equation with an immediate solution of the form
\begin{equation}
\rho_m=\rho_0 a^{-3\gamma}
\label{b13}\end{equation}
where $\rho_0$ is an integration constant.  Alternatively, this solution can be written as
\begin{equation}
\rho_m=\rho_0 a^{-3(1+w)+\varepsilon}
\label{b14}\end{equation}
with $\varepsilon=\frac{2n}{m}(w+1)$.  This states that when $\varepsilon>0$ matter is created and energy is constantly injecting into the matter so that the latter will dilute more slowly compared to its standard evolution $\rho_m \propto a^{-3(w+1)}$.  Similarly, when $\varepsilon<0$ the reverse is true, namely that matter is
annihilated and direction of the energy transfer is outside of the matter system so that the rate of the dilution is faster than the standard one. In this
interacting case, it is shown \cite{bis1} that cosmological set up of the gravitational field equations
(\ref{2}) actually admit a class of power law solutions with
\begin{equation}
m=\frac{2(nw+1)}{3(w+1)}
\label{b9-1}\end{equation}
which indicates accelerating expansion in certain conditions.  \\
Let us investigate some thermodynamic features of the non-minimal
coupling of matter described by (\ref{1}).  A thermodynamic
description of a perfect fluid matter system requires the knowledge
of the particle flux $N^{\alpha}=qu^{\alpha}$ and the entropy flux
$S^{\alpha}=q\sigma u^{\alpha}$ where $q = N/a^3$ and $\sigma = S/N$
are, respectively, the concentration and the specific entropy (per
particle) of the created or annihilated particles. Since the energy
density of matter is given by $\rho_m=qM$ with $M$ being mass of
the particles, the appearance of the extra term in the energy
balance equation (\ref{b10}) means that this extra-change of
$\rho_m$ can be attributed to a change of $q$ or $M$.  Here we
assume that the mass of each matter particle remains constant and
the extra term in the energy balance equation only leads to a change
of the number density $q$. In this case, the equations (\ref{b10})
can be written as
\begin{equation}
\dot{q}+3(w+1)qH=q\Gamma
\label{c15}\end{equation}
where
\begin{equation}
\Gamma\equiv-(w+1)q\frac{\dot{R}}{R}=\varepsilon H=2n(w+1)t^{-1}
\label{c16}\end{equation}
 is the decay rate.  We also assume that the overall energy transfer is an adiabatic processes in which matter particles are continuously created or annihilated while
 the specific entropy per particle remains constant during the whole processes ($\dot{\sigma}=0$) \cite{li}.  This means that
 \begin{equation}
 \frac{\dot{S}}{S}=\frac{\dot{N}}{N}=\Gamma
 \label{c17}\end{equation}
The second law of thermodynamics is fulfilled when $\dot{S}\geq 0$
or, equivalently, $\Gamma\geq 0$.  From $n\propto
a^{-3+\varepsilon}$, we can see that the total number of particles
scales as $N\propto a^{\varepsilon}$, and $\dot{S}\geq 0$ implies
that $\varepsilon\geq0$ in an expanding universe or $\varepsilon<0$
in a shrinking one.  This conclusion can also be drawn by
(\ref{c16}) since one can consider $\Gamma\geq0$ independent of the
sign of $m$. \\The last equality in (\ref{c16}) implies that there
are two cases consistent with
the second law of thermodynamics :\\
First, when $n>0$ and $w+1>0$.  This corresponds to a fluid satisfying weak energy condition.  From (\ref{b9-1}), one can see
that there is an accelerating expansion ($m>1$) for the scale factor for $w>0$ if $n>\frac{3}{2}+\frac{1}{2w}$.  On the other
hand, when $-1<w<0$ there are solutions that are expanding ($m>0$) if $n<\frac{1}{|w|}$ and shrinking ($m<0$)
if $n>\frac{1}{|w|}$. \\
Second, when $n<0$ and $w+1<0$. The equation of state
$w<-1$ corresponds to a phantom fluid.  In this case, exploring
(\ref{b9-1}) reveals that this solution describes a shrinking
universe ($m<0$) which can be changed to an expanding one by
choosing $t<0$.  The solution would be also accelerating $|m|>1$ if
$|n|>\frac{3|w+1|}{2|w|}-\frac{1}{|w|}$. It is
interesting to compare this situation with the standard phantom
cosmology \cite{car}. The latter corresponds to the non-interacting
limit ($\lambda=0$) of (\ref{1}), or $\varepsilon=0$ in (\ref{b14}),
when a perfect fluid with a constant equation of state parameter
$w$ is taken as a matter system.  In this case, the energy
density of the fluid and the scale factor scale as $\rho\propto
a^{-3(w+1)}$ and $a\propto t^{\frac{2}{3(w+1)}}$,
respectively.  For $w < -1$, the energy density grows with time
so that it becomes infinite in the future.
\\ In our case, however, the evolution of energy density is modified
due to the non-minimal interaction with geometry, as described by
(\ref{b14}). The relation indicates that there are two terms in the
exponent that determine evolution of the energy density.  The first
term, which makes the energy density grow with expansion of the
scale factor and the second term which appears as a compensating
term due to the fact that $m<0$.  In fact, in this case
$\varepsilon<0$ and the effect of the interaction is annihilation of
the phantom fluid. The annihilation process can weaken the growing
of the energy density of the phantom or even cease its
increasing and start decreasing if $|\frac{n}{m}|>\frac{3}{2}$.\\
In summary, we have considered a non-minimal coupling of matter
systems with geometry via a function of the Ricci scalar $f_2(R)$
which leads to non-conservation of matter energy-momentum tensor.
Assuming a power-law form for the scale factor and the function
$f_2(R)$, we have solved the (non-)conservation equation in the two
cases $\lambda \alpha R^{n}<<1$ and $\lambda \alpha R^{n}>>1$.  In
the first case, there is nearly no energy transfer between the two
components and matter stress tensor is conserved. In the second
case, however, there is a constant rate of energy transfer. In both
cases there is no extra force in the geodesic equation as the choice
$L_m=p_m$ leads to vanishing of the
first term on the right hand side of the equation (\ref{b7}).  We have already used this approach to investigate non-minimal coupling of a perfect fluid matter system
in a cosmological setting \cite{bis1}.  In that work, our primary interest was to answer the question that whether the accelerating expansion
of the universe could be realized in such an interacting model.  We have shown that the answer can be affirmative if certain conditions are satisfied.  In particular, the parameters $m$ and $n$ are constrained by accelerating expansion and the second
law of thermodynamics.  In the present work, we have considered non-minimal coupling of a phantom fluid.  Attention here is focused on the possibility that
in the interacting case ($\lambda\alpha R^n>>1$) decaying of the phantom avoids the Big Rip singularity.
\\
A thermodynamic description of the non-minimal coupling reveals that
among different possibilities there are only two cases that are
consistent with the second law of thermodynamics; $n>0$,
$w+1>0$ and $n<0$, $w+1<0$.  The latter case is of
particular importance since it considers non-minimal coupling of a
phantom fluid with geometry.  Our analysis indicates that due to the
interaction which actually appears as an annihilation process
($\varepsilon<0$), the energy density of such a fluid decreases with
expansion of the universe if $|n|>\frac{3}{2}|m|$. In this case, the
universe avoids Big Rip singularity in the future.
~~~~~~~~~~~~~~~~~~~~~~~~~~~~~~~~~~~~~~~~~~~~~~~~~~~~~~~~~~~~~~~~~~~~~~~~~~~~~~~~~~~~~~~~~~~~~~~~~~~~~~~~~~~~~~~~~~~~~~~~~~~~~~~~~~~~

\end{document}